# Field Evaporation of Grounded Arsenic Doped Silicon Clusters

Zexiang Deng, Juncong She, Zhibing Li and Weiliang Wang[*]

State Key Laboratory of Optoelectronic Materials and Technologies, School of Physics and Engineering, Sun Yat-sen University, Guangzhou 510275, China

Email: wangwl2@mail.sysu.edu.cn

## Abstract

We have investigated field evaporation of grounded arsenic (As) doped silicon (Si) clusters consist of 52 atoms with density functional theory to mimic Si nano structures of hundreds of nanometers long standing on a substrate. Six cluster structures with different As doping concentrations and dopant locations are studied. The critical evaporation electric fields are found to be lower for clusters with higher doping concentrations and doping sites closer to the surface. We attribute the difference to the difference in binding energies corresponding to the different As-doping concentrations and to the doping locations. Our theoretical study could shed light on the stability of nano apexs under high electric field.

## I. Introduction

Silicon tip emitters have been investigated for applications in nano electronic devices in the last few decades. The silicon based nano materials have gained a lot of attentions for both the applications and the basic understanding; because of their well-understood electronic properties and the fact that silicon based field emission devices have the possibility of integration with various monolithic circuits [1]. Recently, many research groups have great interests in synthesis of well-organized nano materials, and have developed the synthesis of high quality silicon nano structures including physical evaporation, laser ablation and chemical vapor deposition method. Among them, silicon nanowires[2,3] have been widely studied, such as the field emission performance. It has been demonstrated that silicon nanowires have shown excellent field emission performance, and under the strong field condition, silicon apex distortions appear at the top of silicon nanowires [4]. Furthermore, the atom probe tomography has been used to study the structure of the apex[7,8]. As the distance between nano structures decreases, the electric field near nano apexs gets higher and higher. It is essential to understand the properties of field evaporation [5, 6, 19, 20, 21].

It is demonstrated that the extent of deformation at the silicon apex depends on the compositions [4]. The present paper aims to investigate the electric field evaporation of grounded As doped Si cluster to mimic a Si nano structure standing on a Si substrate. Since the system is in nano scale, the quantum effect is essential, which requires us using the quantum mechanical many-body theory. The DFT(density functional theory)[ 9,10] is helpful solving such many-body problems and providing explicit electronic properties of nano system. Because the whole system is in equilibrium as an approximation, the statistical mechanical grand canonical ensemble theory is applied[11,12] to account the exchange of electrons between the Si cluster and the substrate.

Sec. II shows the detailed method of calculation and Sec .III is the calculation results and discussions, while the conclusions are given in Sec. IV.

## II. Calculation Method

Following the floating sphere model [16], we replace a nano structure standing on a substrate with a grounded nano cluster. We grounded the nano cluster by assuming it in equilibrium with the substrate. We investigate the electric field evaporation of Si cluster with density functional theory implemented in the Dmol3 code [13,14]. The PWC [15] of the local density approximation (LDA) is used for the exchange-correlation functional and the double numerical plus d-functions (DND) basis set is applied for all electrons. The energy and displacement convergence tolerance in geometry optimization is $2.0 \times 10^{-5}$ *Ha* and $5.0 \times 10^{-3}$ angstroms respectively.

We have picked six typical structures to investigate the effect of doping As atoms on the threshold electric field of evaporation. In Fig.1 (a), there is no doping atoms, Fig.1 (b)~(c), Fig.1 (d)~(f) has one and two doping atoms respectively.

In the present work, Si cluster is grounded, i.e. assuming that Si cluster is in equilibrium with the substrate, we employ grand canonical ensemble to account the exchange of electrons between Si cluster and the substrate. Since we just consider the ideal condition, the possible barrier caused by interface contact, i.e. *Schottky* barrier is neglected. The Fermi level in the cluster is equal to the Fermi level of substrate (in this article, the substrate is crystalline silicon with Fermi level -5 *eV*) while there is no external applied field. The applied field *F* will lower the energy levels in Si cluster by *FL* or raise the Fermi level in Si cluster by *FL* in other words, where *L* is the distance from Si cluster to silicon substrate.

The atomic structure of Si cluster is optimized without external applied electric field, and then is fixed while we determine the number of electrons *N* in grounded Si cluster under applied electric field *F*. We calculate the energy of the system $E_{N,s}$ with given *N*. Following the grand canonical ensemble theory, the possibility of the state

$$\rho_{N,s} = \frac{\exp\{\beta[(\mu + FL)N - E_{N,s}]\}}{\sum_{N,s} \exp\{\beta[(\mu + FL)N - E_{N,s}]\}},$$

where $\beta = 1/k_B T$, $\mu$ the chemical potential of the crystalline silicon substrate, $k_B$ the Boltzmann constant, *T* the temperature of substrates.

The atomic structure of Si cluster with the most probable electron number $N_p$ is relaxed under applied electric field *F*, to find out whether the structure will be broken. Then we can determine the threshold electric field of evaporation.

## III. Results and Discussions

The critical external field $F_C$ is found to be inversely proportional to distance *L* between Si cluster and the substrate for all six Si clusters (Fig. 2). With more As atoms doped or the As atom doped closer to the apex, the $1/F_C - L$ slope *k* is larger. i.e. The slope *k* follow hierarchy: *k*(a) < *k*(b)<*k*(c) <*k*(d)<*k*(f)<*k*(e), where *k*(a) is the $1/F_C - L$ slope for structure (a) and *etc.*.

In general, the external field *F* we applied is called *macroscopic field,* which is different from the local field $F_L$, near the apex, that determines the breaking down of the apex.

According to the floating sphere model[16]

$$1/F = (2.5 + L/r) / F_L$$

where *r* is the radius of the apex. *1/F* is a linear function of *L* which agrees with Fig.2. C. J. Edgcombe *et al.* [18] calculated the field enhancement factor $\gamma$ for various geometries and sizes of CNTs by means of the finite element method which demonstrated the similar *1/F~L* relation. However, the difference of slope for different Si cluster with the same radius *r* could not explained by the 'floating sphere' model [16], which should be explained by a more accurate model that takes the detailed electronic properties of Si cluster into account.

We have calculated the binding energies of the six structures. The binding energy is defined as $E_b = E_{atoms} - E_{cluster}$, where $E_{atoms}$ is the total energy of isolated atoms and $E_{cluster}$ is the energy of cluster. Table I. lists the $1/F_C - L$ slope *k* and binding energies corresponding for the six structures. Thomas Bschel *et al.* found that the binding energies for neutral silicon clusters with sizes in the range of 25~70 atoms are around 4 *eV* per atom [17], which is close to our results in third row of table I. The binding energies follow the reverse hierarchy as slope *k*: $E_b$(a)>$E_b$(b)>$E_b$(c)>$E_b$(d)>$E_b$(f)>$E_b$(e). This explains the hierarchy of slope *k*: the critical external field $F_C$ is a decreasing function of $E_b$ ; and the slope *k* is inversely proportional to $F_C$. Therefore, the slope *k* should follow the reverse hierarchy as the binding energy. The structure dependence of the binding energies can be explained as a consequence of different cluster electron densities. The doped Si clusters have higher electron densities than those without or less doping atoms. With higher electron densities, the stronger coulomb repulsion may reduce the binding energies. Therefore $E_b$(a) > [$E_b$(b) and $E_b$(c)] > [$E_b$(d), $E_b$(f) and

$E_b$(e)]. More explicitly, the doping atoms locate closer to the surface (especially the apex) creates higher local electron density. Fig.3 shows the electron density differences between structure (b~f) and structure(a). In one doping atom case, the doping atom in structure (c) has larger Mulliken charge (0.552 *e*) than that of structure (b) (0.539 *e*), which means higher local electron density and lower binding energy. Therefore $E_b$(b)>$E_b$(c). In two doping atoms case, the Mulliken charge is not the predominant factor, but the two doping site distance. The distance between the two doping atoms in structure (d, e, f) is $D$(e) < $D$(f) < $D$(d). The smaller distance $D$ results in weaker Coulomb repulsion . Therefore $E_b$(d)>$E_b$(f)>$E_b$(e).

## IV. Conclusions

The electric field evaporation of grounded As doped Si cluster has been investigated with density functional theory. The Si cluster system is assumed in equilibrium with substrate and follows the grand canonical ensemble theory.

The critical field $F_C$ is found to be inversely proportional to the distance $L$ between the Si cluster and the substrate. The slope $k$ of $1/F_C - L$ curve increases with the binding energy of the cluster which decreases with the doping concentrations. For the clusters with the same doping concentration, closer the doping atoms are to the apex, smaller the $k$ is. Which means higher doping concentration or doping atoms closer to the apex will reduce the critical evaporation electric field.

ACKNOWLEDGEMENTS
The project was supported by the National Basic Research Program of China (2013CB933601), the National Natural Science Foundation of China (Grant No. 11274393, 11104358), the Fundamental Research Funds for the Central Universities (No. 13lgpy34) and the high-performance grid computing platform of Sun Yat-sen University.

**Figures**

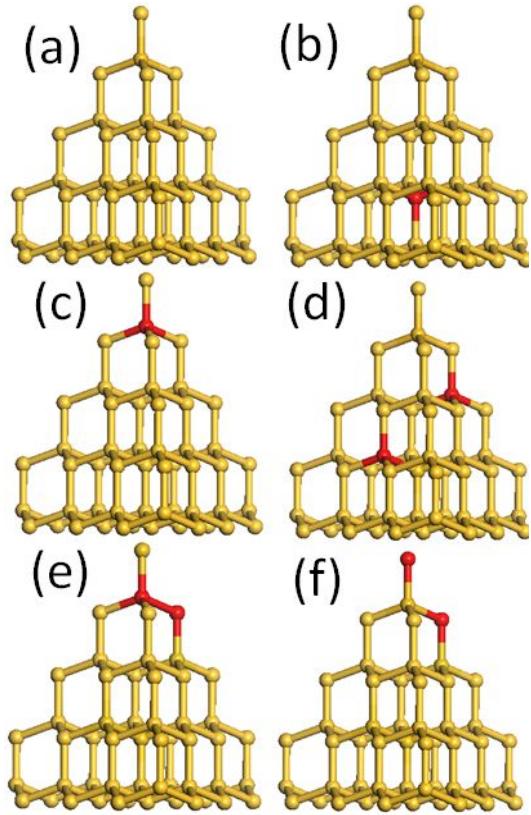

Fig.1. Six typical silicon clusters. The yellow ball refers to silicon atom, while the red one means arsenic atom.

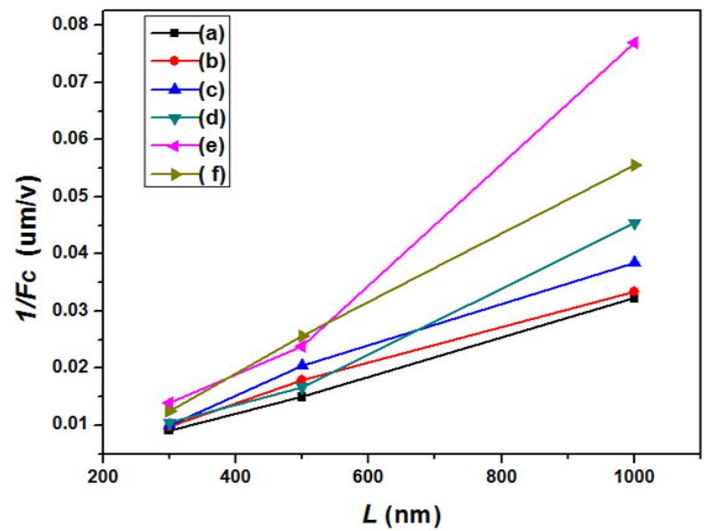

Fig.2. $1/F_C - L$ cure, where $F_C$ is the critical external electric field, $L$ is the distance between the Si cluster and the substrate. (a-f) correspond to the six structures in Fig. 1. Lines are to guide the eyes.

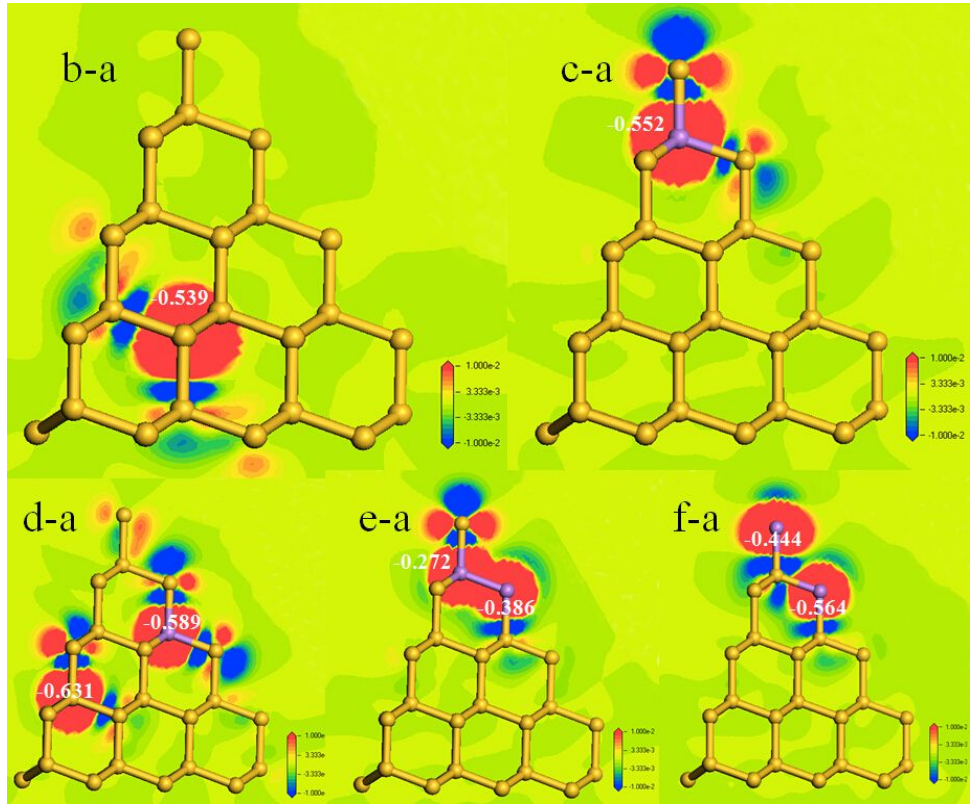

Fig.3 The electron density differences of structure (b~f) and structure(a), which is corresponding to the structures in Fig.1. b-a is the density of structure(b) subtracts that of structure(a) and *etc.*. The red region indicates more electrons of the current structure compared with structure(a), while the blue one has the opposite meaning. And the white digitals are the Mulliken charges of doping atoms (As).

Table I. The slope $k$ and the binding energies corresponding to the six structures in Fig. 1.

|  | a | b | c | d | e | f |
|---|---|---|---|---|---|---|
| Slope $k$ (1/V) | 0.033 | 0.034 | 0.041 | 0.052 | 0.079 | 0.061 |
| Binding Energy (eV) | 215.501 | 214.210 | 210.670 | 210.144 | 209.866 | 210.042 |
| BE per Atom (eV) | 4.144 | 4.119 | 4.051 | 4.041 | 4.036 | 4.039 |